\documentstyle[multicol,prl,aps,epsf] {revtex}
\begin{document}
\bibliographystyle{prlsty}
\draft
\title{Spin-Glass and Chiral-Glass Transitions in a 
$\pm J$ Heisenberg Spin-Glass Model in Three Dimensions}
\author{Tota Nakamura and Shin-ichi Endoh}
\address{Department of Applied Physics, Tohoku University,
         Aoba-yama 08, Sendai, 980-8579, Japan }
\date{\today}
\maketitle
\begin{abstract}
The three-dimensional $\pm J$ Heisenberg spin-glass model is investigated by
the non-equilibrium relaxation method from the paramagnetic state.
Finite-size effects in the non-equilibrium relaxation are analyzed, and
the relaxation functions of the spin-glass susceptibility and the chiral-glass
susceptibility in the infinite-size system are obtained.
The finite-time scaling analysis gives
the spin-glass transition at 
$T_{\rm sg}/J=0.21 _{-0.02}^{+0.01}$
and the chiral-glass transition at 
$T_{\rm cg}/J=0.22 _{-0.03}^{+0.01}$.
The results suggest that both transitions occur simultaneously.
The critical exponent of the spin-glass susceptibility is estimated as
$\gamma_{\rm sg}= 1.7 \pm 0.3$, which makes an agreement with the
experiments of the insulating and the canonical spin-glass materials.
\end  {abstract}


\begin{multicols}{2}
\narrowtext

It has been an unsettled issue whether a short-ranged spin-glass model can 
explain the spin-glass transition observed in real materials.\cite{youngref}
The existence of the spin-glass transition in the Ising model 
has been widely recognized:
various types of investigations 
\cite{isings,kawashima}
gave consistent values of the transition temperature.
The recent large-scale simulation study \cite{kawashima} evaluated the 
critical exponents which quantitatively agree with the measurements of the 
Ising-type spin-glass material, Fe$_{0.5}$Mn$_{0.5}$TiO$_3$.%
\cite{gunnarsson}
However,
the spins and the magnetic interactions of many spin-glass materials
are mostly isotropic,
and the anisotropy effect can be regarded as a
secondary effect.
Therefore,
the Heisenberg model with random interactions is considered 
to be the simplest theoretical model.
The problem is whether this model exhibits the
spin-glass transition or not.
Previous numerical investigations
suggest that the finite-temperature transition does not occur.%
\cite{mcmillan,olive86,kawamura92,kawamura95,kawamura98,%
hukushima2000,matsubara91}

In order to solve this contradiction, Kawamura \cite{kawamura92}
introduced the chirality mechanism.
The chirality is a variable whose absolute value corresponds to a volume 
of a parallelepiped spanned by three spins, 
and whose sign represents the handedness of the spin structure.
In the chirality mechanism,
the chirality is considered to freeze randomly at a finite temperature
without the conventional spin-glass order.
The experimentally-observed spin-glass anomaly is considered as
an outcome of 
a mixture of the chirality with the spin by the weak magnetic anisotropies.
The existence of the chiral-glass transition has been confirmed by 
several numerical 
analyses.\cite{kawamura92,kawamura95,kawamura98,hukushima2000}
However,
it was pointed out recently by one of the present authors (S. E.) that 
the finite-temperature spin-glass transition may exist
by examining the stiffness of the system when all the spins on a surface are
rotated and reversed, respectively.\cite{matsu2000,endo2001}
The same evidence was also reported by a dynamical
Monte Carlo simulation which removes the global spin rotation.\cite{matsu2001}

In this Letter, we focus on the existence or the non-existence of the
spin-glass and the chiral-glass transitions.
By employing the non-equilibrium relaxation method,
we have revealed that both transitions may occur simultaneously
at a finite temperature $T/J \simeq 0.21$.
This result possibly settles the discrepancies between the experiments
and the theories in the Heisenberg spin glass in three dimensions.

In the simulational studies of the phase transitions, one obtains 
equilibrium quantities in finite-size systems, and the finite-size scaling
is utilized to extract the thermodynamic properties from them.
However, 
the dynamics of the simulations become very slow due to the
randomness and frustration in the spin-glass models,
which makes it hard to realize the equilibrium state in large systems.
The system size accessible by the equilibrium simulations is restricted
to linear sizes $L \sim 30$ up to now.
In this situation, reliability of the finite-size scaling might become
doubtful, which causes controversial conclusions of the existence
\cite{matsu2001pre} or the nonexistence \cite{olive86} 
of the spin-glass transition 
depending on the sizes and the temperatures of the data used.
In this Letter, we follow a completely alternative approach to the 
thermodynamic limit, i.e., 
we observe the relaxation of the infinite-size system to the equilibrium
state.
In order to extract the equilibrium properties, the finite-time scaling 
analysis is utilized instead of the conventional finite-size scaling.
This approach is known as the non-equilibrium relaxation (NER) method.
\cite{stauffer92,ito93,ito99,sadic84,huse89,blundell92,oz-it-og,oz01,shira01}
Actually, we prepare a very large lattice with the paramagnetic state (
perfectly random spin configuration), and observe the relaxation of
the susceptibility.
The simulations are stopped before the finite-size effect appears.
Time dependences of the susceptibility 
diverge algebraically as $\chi(t) \sim t^{\gamma /z\nu}$ 
\cite{ito99,sadic84,huse89,blundell92}
in the 
critical region, while it converges to a finite value in the 
paramagnetic region.
We detect the phase transition by this difference.
Another advantage of this strategy is an absence of a trivial 
exponential decay due to the global spin rotation.
Since the spin autocorrelation function is essentially 
a one-body function about the spin, 
one may encounter this exponential decay 
even in the ordered phase after a time
scale of the global spin rotation.\cite{matsu2001}

We consider the $\pm J$ Heisenberg spin-glass model on a simple cubic lattice
of $N=L\times L\times (L+1)$
under the skew periodic boundary conditions,
$
{\cal H} = -\sum_{\langle i, j\rangle}
 J_{ij} \mbox{\boldmath $S$}_{i}\cdot 
                        \mbox{\boldmath $S$}_{j}.
$
The exchange interactions $J_{ij}$ take a value of $+J$ or $-J$ with the 
same probability, and $\mbox{\boldmath $S$}_{i}$ is a three-component vector
spin with $|\mbox{\boldmath $S$}_{i}|=1$.
The summation runs over all the nearest-neighbor spin pairs.
We evaluate the spin-glass susceptibility, $\chi_{\rm sg}$, 
and the chiral-glass susceptibility, $\chi_{\rm cg}$,
through the following relations.
\begin{eqnarray}
\chi_{\rm sg}&=&\frac{1}{N}\sum_{i,j} \left[\langle 
\mbox{\boldmath $S$}_{i}\cdot \mbox{\boldmath $S$}_{j}
\rangle^2 \right]_{\rm c}
=
N\sum_{\mu,\nu} \left[\langle q^2_{\mu,\nu}\rangle \right]_{\rm c},
\\
\chi_{\rm cg}&=&\frac{1}{3N} [\langle (
\sum_{i, \mu}
C_{i, \mu}^{(a)}
C_{i, \mu}^{(b)})^2\rangle ]_{\rm c},
\end  {eqnarray}
where
$q_{\mu, \nu}=(1/N)\sum_i S_{i, \mu}^{(a)} S_{i, \nu}^{(b)}$
is a replica overlap between the $\mu$ component of a spin $i$ 
on a replica $(a)$, $S_{i, \mu}^{(a)}$,
and the $\nu$ component of the spin on a replica $(b)$, $S_{i, \nu}^{(b)}$.
The bracket $[\cdots ]_{\rm c}$ denotes the configurational average, while 
$\langle \cdots \rangle$ denotes the thermal average.
The indices, $\mu$ and $\nu$, stand for three components of spins:
$ x, y,$ and $ z$.
The local chirality $C_{i, \mu}^{(a)}$ is defined by
three neighboring spins as
$
C_{i, \mu}^{(a)}=
\mbox{\boldmath $S$}_{i+\hat{\mbox{\boldmath $e$}}_{\mu}}^{(a)}
\cdot
(
\mbox{\boldmath $S$}_{i}^{(a)}
\times
\mbox{\boldmath $S$}_{i-\hat{\mbox{\boldmath $e$}}_{\mu}}^{(a)}
),
$
and $\hat{\mbox{\boldmath $e$}}_{\mu}$ denotes a unit lattice vector along
the $\mu$ axis.

We use the single-spin-flip algorithm with the heat-bath probability.
\cite{olive86}
For a given random bond configuration,
eight replicas are prepared with different random initial spin configurations.
Each replica is updated in parallel by a different random number sequence, and 
twenty eight replica-overlaps are calculated and averaged.
Note that these data are correlated with each other, 
and only seven among them are independent.
Then, we take averages of $\chi_{\rm sg}(t)$ and $\chi_{\rm cg}(t)$ 
over the bond distributions.
Typical numbers of the bond configurations are tens to
thousands depending on the sizes, the temperatures, and the time ranges.

From the non-equilibrium relaxation functions of $\chi_{\rm sg}$ and
$\chi_{\rm cg}$,
we extract the equilibrium properties
and estimate the transition temperature by the finite-time scaling analysis,
\cite{oz-it-og,shira01}
which is a direct interpretation of the conventional finite-size scaling
through the relation, $\tau \sim \xi^z$.
We determine the correlation time $\tau(T)$ at each temperature and
a combined exponent
$\lambda (\equiv \gamma/z\nu)$
so that the scaled functions, $\chi(t) t^{-\lambda}$, 
for various temperatures fall onto a
single curve when plotted against $t/\tau(T)$.
Then, we extrapolate $\tau(T)$ with an assumption
that they diverge algebraically
when the temperature approaches the transition temperature $T_{\rm c}$ as
$
\tau(T) \propto (T-T_{\rm c})^{-z\nu}.
$
We obtain the most probable estimate for $T_{\rm c}$ 
by the least-square fitting of this form.

\begin{figure}
 \epsfxsize = 8.0cm
 \epsffile{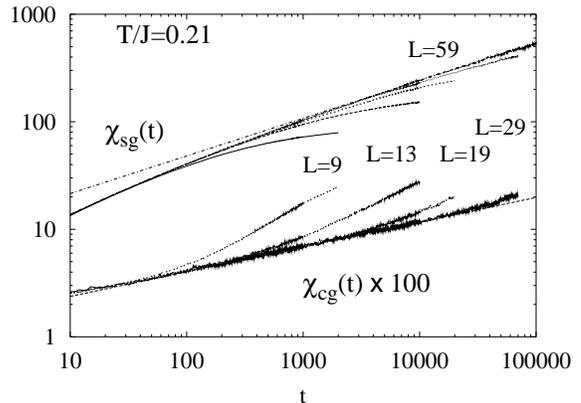}
 \caption {
The size dependences of $\chi_{\rm sg}(t)$ (up to $L=59$) 
and $\chi_{\rm cg}(t) $ multiplied by 100 (up to $L=29$)
at the temperature $T/J=0.21$.
After the initial relaxation,
$\chi_{\rm sg}(t)$ and $\chi_{\rm cg}(t)$ converge to critical behaviors 
$t^{\lambda}$ (dotted straight lines) with $\lambda=0.35$(sg),
and $\lambda=0.23$(cg).
These exponents coincide with the estimates obtained by the 
finite-time scaling.
  \label{fig:chisc}
          }
\end  {figure}

In the finite-time scaling, it is crucial to use the data free from the 
finite-size effect.
We estimate a characteristic time 
that the finite-size effect appears by changing the
system size from $L=9$ to $L=59$ at each temperature.
The typical example is shown in Fig. \ref{fig:chisc}.
The temperature is within our estimate of the spin-glass transition 
temperature obtained afterwards.
The finite-size effect of $L=19$ data already appears at $t=1000$ steps, after
which a long relaxation to the finite-size equilibrium state is waiting.
This ``finite-size'' characteristic time is roughly scaled as 
$\tau \sim L^z$ with the dynamic exponent $z\sim 5$.
By this relation, we confirm that the relaxation of $L=59$ before 
$t=100000$ steps is not influenced by the finite-size effect,
and thus can be considered as that of the infinite system.
At each temperature, we check this time scale and use only the data 
with the approval.

In this figure, we notice that the appearance of the finite-size effect is
in opposite ways between the spin-glass susceptibility and the chiral-glass 
susceptibility.
The former is always underestimated, while the latter is overestimated 
showing upward bending.
This tendency continues to the equilibrium values;
the chiral-glass susceptibility always decreases 
as the system size is enlarged.
This size dependence is quite strange as the susceptibility, 
but it can be understood by considering a definition of the chirality.
Since the chirality corresponds to a volume of a parallelepiped
spanned by three neighboring spins, it becomes larger when the 
relative angles between the neighboring spins are enhanced by the
finiteness of the lattices.
Thus,
we must pay much attention to the finite-size effects in the relaxation
of the chirality.
For example, 
at paramagnetic temperatures,
the chiral-glass susceptibility converges to a finite value 
in the large size limit,
while it shows an upward bending in the small size limit.
There exists a crossover size in which the non-equilibrium relaxation 
looks algebraically divergent as if it exhibits the criticality.
This crossover size is considered to correspond to the chiral correlation
length.
If one tries to detect $T_{\rm cg}$ solely by finding the 
algebraic relaxation behavior,
he may be misled to a wrong temperature where
$\xi = L$ instead of $\xi = \infty$.
Therefore, 
the chiral-glass transition temperature tends to be overestimated,
and the spin-glass transition temperature is to be underestimated, 
if the system size is insufficient.

As shown in Fig. \ref{fig:chisc},
the initial relaxation of the spin-glass susceptibility continues until
$t \sim 10^3$ steps, then it exhibits the critical relaxation.
In the finite-time scaling analysis, we must discard this initial relaxation
data.
It is also noticed that the finite-size effects of the small sizes
($L \le 19$) 
appear before the initial relaxation ends.
The critical phenomena may not be extracted from the
equilibrium states of such small systems.

\begin{figure}[h]
    \epsfxsize = 8.0cm
\epsffile{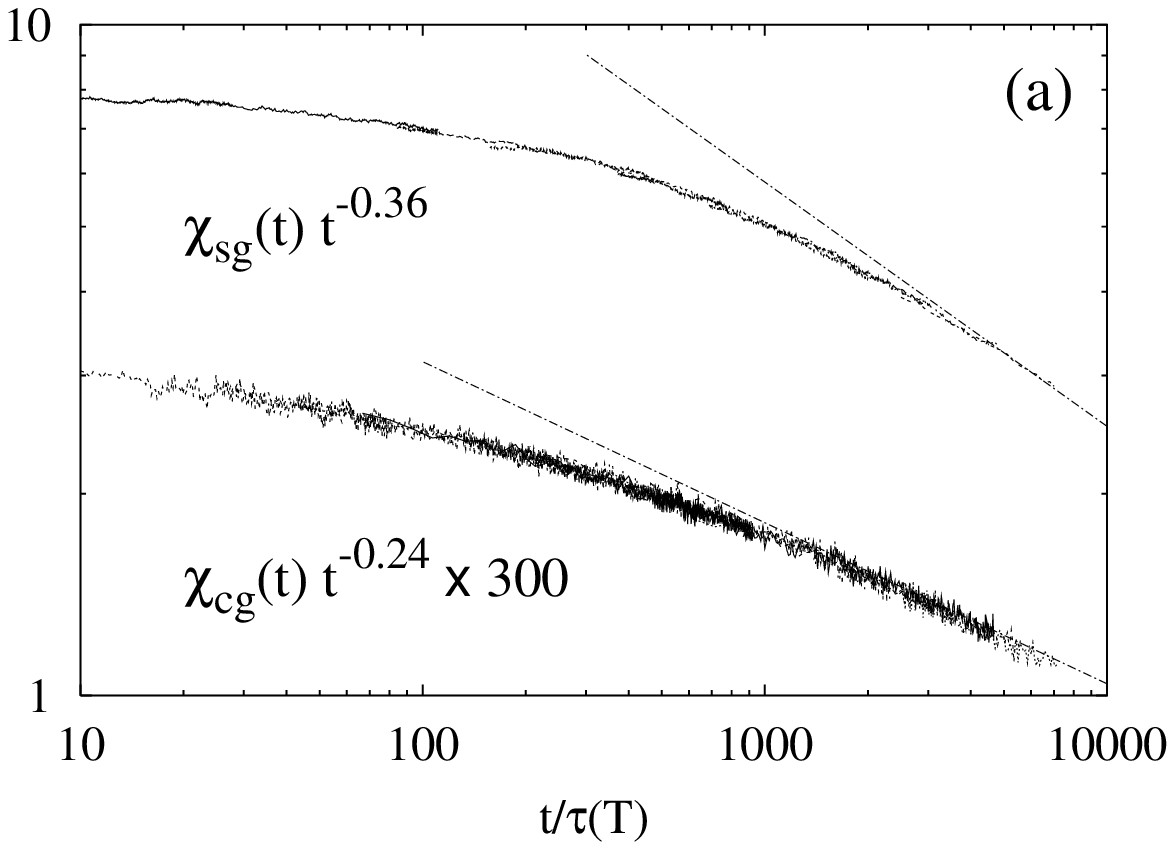}

    \epsfxsize = 8.0cm
\epsffile{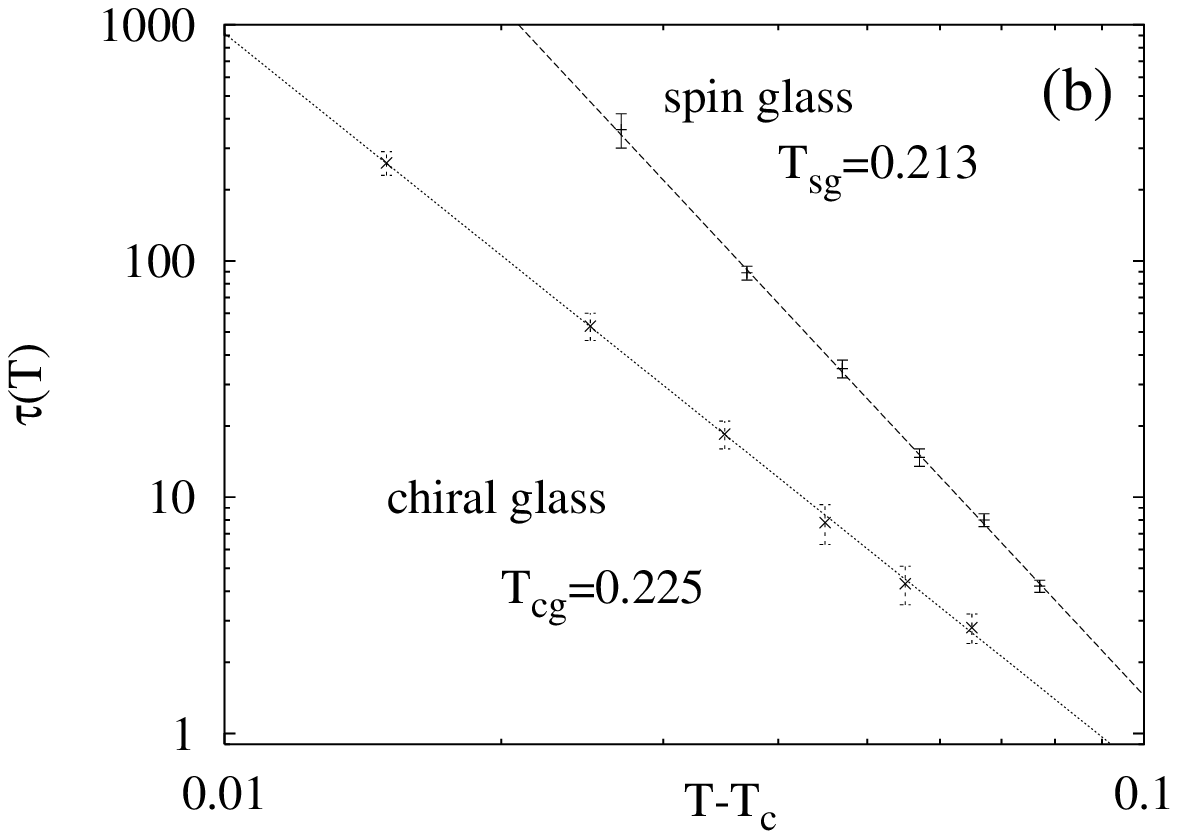}
 \caption {
(a)
  The finite-time scaling of $\chi_{\rm sg}(t)$ with 
$\lambda=\gamma/z\nu=0.36$
and $\chi_{\rm cg}(t)$ (multiplied by 300) with $\lambda=0.24$. 
The $t^{-\lambda}$ are depicted by straight lines.
The data of $L=59$ at
$T/J=$0.32, 0.29, 0.28, 0.27, 0.26, 0.25, and 0.24 are plotted.
  \label{fig:sca}
(b)
  Extrapolations of the correlation time $\tau(T)$ of
$\chi_{\rm sg}$ and $\chi_{\rm cg}$ obtained by the scaling. 
The transition temperatures are estimated so as to minimize the least-square
errors.
Six temperatures ranging $0.24 \le T/J \le 0.29$ are used.
  \label{fig:tau}
          }
\end  {figure}

Results of the finite-time scaling of the spin-glass susceptibility 
and the chiral-glass susceptibility are shown in 
Fig. \ref{fig:sca}.
The combined exponent $\lambda$
and 
the correlation time $\tau(T)$ 
are estimated
by using the data of $L=59$ until $t=70000$ steps at seven temperatures
ranging $ 0.24 \le T/J \le 0.32$.
The data in the initial relaxation are discarded.
The scaling function converges
to a $t^{-\lambda}$ line in the equilibrium region ($t/\tau(T) \to \infty$),
where the susceptibility converges to an equilibrium value in the 
infinite system.
On the other hand, it becomes a flat line in the non-equilibrium limit,
 $t/\tau(T) \to 0$, 
since the susceptibility behaves as $\chi(t)\sim t^{\lambda}$ in the 
critical region.
Therefore, this curve interpolates the
non-equilibrium and the equilibrium limits. 

The correlation time is extrapolated by $(T-T_{\rm c})^{-z\nu}$,
and we obtain the transition temperature that gives the
best least-square fitting.
The exponent is estimated to be $z\nu= 4.7\pm 0.8$.
Combining this value with $\lambda=\gamma/z\nu= 0.36\pm 0.02$,
the exponent for the spin-glass susceptibility is estimated.
\[
T_{\rm sg}/J=0.21 _{-0.02}^{+0.01} ,~~~
\gamma_{\rm sg} = 1.7 \pm 0.3.
\]
This transition temperature agrees well with that obtained by
the defect free-energy method,\cite{endo2001} and that obtained by
the dynamical simulations.\cite{matsu2001}
The exponent is consistent to that 
of the non-linear susceptibility in
real spin-glass compounds:
$\gamma=2.2\pm 0.1$ (CuMn \& AgMn)\cite{sgmate1} and
$\gamma=2.3\pm 0.4$
(Cd Cr$_{2\times 0.85}$In$_{2\times 0.85}$S$_4$).\cite{sgmate2}
This may suggest that the present theoretical model 
is relevant to the spin-glass materials.
Estimates and the comparison of other exponents are left for a 
future investigation.
It requires more detailed and accurate calculations of the Binder ratio,
which diverges in the critical region as $t^{d/z}$.
Preliminary calculations of the Binder ratio, however, suggest that the 
size effects and the sample dependences are much severer than those
of the susceptibility.
The algebraic divergence is only observed in lattices larger than 
$L=19$, below which the finite-size effects appear in the initial
relaxation.
The details will be reported elsewhere.

Analyses of the chiral-glass transition are done in the same procedure.
Note that the amplitude of $\chi_{\rm cg}$ is smaller 
than that of $\chi_{\rm sg}$ about two digits, which causes rather
large numerical errors.
A possible value of $\lambda$ ranges as $\lambda = 0.23\pm 0.04$.
The transition temperature and its exponent are obtained as
\[
T_{\rm cg}/J=0.22 _{-0.04}^{+0.01}, ~~~
\gamma_{\rm cg} = 1.0 \pm 0.5.
\]
Note that this value is consistent with the estimate obtained by the
equilibrium simulations which gives $T_{\rm cg}\sim 0.21$.
\cite{hukuprivate}
The value of $\gamma_{\rm cg}$ does not contradict with that of the
model with the Gaussian bond distributions.\cite{hukushima2000}
The exponents $\lambda$ for $\chi_{\rm sg}$ and $\chi_{\rm cg}$ 
estimated in the scaling also coincide with the slope of the 
relaxation functions at the transition temperature in Fig. \ref{fig:chisc}.
This evidence supports the validity of the present analysis.
By the results presented in this Letter,
it is conjectured that the spin-glass transition and 
the chiral-glass transition occur at the same temperature.
Since the chirality must freeze if the spin freezes, the transition
is possibly driven by the spin degrees of freedom.

We have observed the simultaneous finite-temperature spin-glass transition 
and the chiral-glass transition in the 
$\pm J$ Heisenberg model in three dimensions.
This was made possible by observing the non-equilibrium relaxation of
the infinite-size system started from the paramagnetic state.
We consider the inconsistency of the present results with those of the
previous investigations %
\cite{mcmillan,olive86,kawamura92,kawamura95,kawamura98,hukushima2000,%
matsubara91}
can be explained by the differences of the lattice size and the 
temperature range investigated.
Generally,
one expects to observe the critical behavior
as long as the lattice size is comparable with or larger than the correlation 
length, $L \ge \xi$.
Contrary,
when the correlation length exceeds the size, $L \ll \xi$, 
a very large but a finite cluster swallows the whole system, 
and only a trivial size effect is observed.
As the temperature is lowered from the paramagnetic phase,
the correlation length exceeds a lattice size at a
temperature above the transition temperature.
If one mixes the data above and below this crossover temperature,
the finite-size scaling analysis may mislead to a wrong conclusion.
Generally, frustration causes a longer correlation length, and
the size of the system is sometimes essential to detect the
phase transition particularly in frustrated systems.\cite{shira01}
In the present non-equilibrium relaxation analysis,
we have used only data of the temperatures and the time ranges 
where they can be considered as the infinite system.
The time range is limited within $10^5$ steps, however, it is 
enough to equilibrate the system at higher temperatures, $T/J\sim 0.3$, 
where we start the scaling analysis.
We gradually lower the temperature so that the scaled relaxation functions
overlap with each other to ride on a single curve.
At lower temperatures, $T/J\sim 0.24$, $10^5$ steps correspond to a 
time scale that the finiteness of the correlation length begins to appear,
i.e., the relaxation function begins to bend.
In order to see a longer relaxation, we need to enlarge the system size 
to eliminate the size effect.
This must become much more time-consuming and we leave it for the future study.
However, it should be noticed that the finite-size effect appears quite
early in the relaxation process as shown in Fig. \ref{fig:chisc}, 
and what is slow is the relaxation to the equilibrium state 
after this ``finite-size'' characteristic time scale.
The relaxation of the infinite system might be simple:
after an initial relaxation of $10^3$ steps, the critical relaxation 
continues to infinity.
As for the relation between the model and real spin-glass materials,
the critical exponent obtained by the scaling, $\gamma_{\rm sg}\sim 1.7$,
agrees with the experiments.
The authors would like to thank Professor Fumitaka Matsubara
for guiding them to this field.
The author T.N. would like to thank Dr. Y. Ozeki, Dr. K. Hukushima,
and Professor H. Takayama for fruitful discussions and comments.
Use of a random number generator RNDTIK programmed by Professor 
N. Ito and Professor Y. Kanada is gratefully acknowledged.
The computation has been done partly at
the Supercomputer Center, Institute for Solid
State Physics, University of Tokyo.

\end{multicols}


\begin{thebibliography}{99}
\bibitem{youngref}
  For a review,
K. Binder and A. P. Young, Rev. Mod. Phys. {\bf 58}, 801 (1986);
J. A. Mydosh, {\it Spin Glasses} (Taylor $\&$ Francis, London, 1993);
{\it Spin Glasses and Random Fields}, edited by A. P. Young 
(World Scientific, Singapore, 1997).

 \bibitem{isings}
 A. J. Bray, and M. A. Moore, J. Phys. C {\bf 17}, L463 (1984);
 R. N. Bhatt, and A. P. Young, Phys. Rev. Lett. {\bf 54}, 924 (1985);
 A. T. Ogielski, and I. Morgenstern, Phys. Rev. Lett. {\bf 54}, 928 (1985);
 R. Singh and S. Chakravarty, J. Appl. Phys. {\bf 61}, 4095 (1987).

\bibitem{kawashima}
N. Kawashima and A. P. Young, Phys. Rev. B {\bf 53}, R484 (1996).

\bibitem{gunnarsson}
K. Gunnarsson, P. Svedlindh, P. Nordblad, L. Lundgren, H. Aruga, and A. Ito,
Phys. Rev. B{\bf 43}, 8199 (1991).

\bibitem{mcmillan}
W. L. McMillan, Phys. Rev. B {\bf 31}, 342 (1985).

\bibitem{olive86}
J. A. Olive, A. P. Young, and D. Sherrington, 
Phys. Rev. B {\bf 34}, 6341(1986).

\bibitem{kawamura92}
H. Kawamura, Phys. Rev. Lett. {\bf 68}, 3785 (1992).

\bibitem{kawamura95}
H. Kawamura, J. Phys. Soc. Jpn. {\bf 64}, 26(1995).

\bibitem{kawamura98}
H. Kawamura, Phys. Rev. Lett. {\bf 80}, 5421(1998).

\bibitem{hukushima2000}
K. Hukushima, and H. Kawamura, Phys. Rev. E {\bf 61}, R1008 (2000).

\bibitem{matsubara91}
F. Matsubara, T. Iyota, and S. Inawashiro, Phys. Rev. Lett. {\bf 67},
1458(1991).

\bibitem{matsu2000}
F. Matsubara, S. Endoh, and T. Shirakura, 
J. Phys. Soc. Jpn. {\bf 69}, 1927 (2000).

\bibitem{endo2001}
S. Endoh, F. Matsubara, and T. Shirakura, 
J. Phys. Soc. Jpn. {\bf 70}, 1543 (2001).

\bibitem{matsu2001}
F. Matsubara, and T. Shirakura, and S. Endoh, Phys. Rev. B {\bf 64},
092412 (2001).

\bibitem{matsu2001pre}
F. Matsubara, and T. Shirakura, and S. Endoh, cond-mat/0011218.

\bibitem{stauffer92}
D. Stauffer, Physica A {\bf 186}, 197 (1992).

\bibitem{ito93}
N. Ito, Physica A{\bf 196}, 591 (1993).

\bibitem{ito99}
N. Ito and Y. Ozeki, Intern. J. Mod. Phys. {\bf 10}, 1495 (1999).

\bibitem{sadic84}
A. Sadic, and K. Binder, J. Stat. Phys. {\bf 35}, 517 (1984).

\bibitem{huse89}
D. A. Huse, Phys. Rev. B {\bf 40}, 304 (1989).

\bibitem{blundell92}
R. E. Blundell, K. Humayun, and A. J. Bray, J. Phys. A {\bf 25}, L733 (1992).

\bibitem{oz-it-og}
Y. Ozeki, N. Ito, and K. Ogawa, ISSP Supercomputer Center Activity Report
1999, 37 (The University of Tokyo, 2000);
Y. Ozeki, K. Ogawa, and N. Ito, in preparation.

\bibitem{oz01}
Y. Ozeki and N. Ito, Phys. Rev. B {\bf 64}, 024416 (2001).

\bibitem{shira01}
T. Shirahata and T. Nakamura, cond-mat/0108414.

\bibitem{sgmate1}
N. de Courtenary, H. Bouchiat, H. Hurdequint, and A. Fert,
J. Physique {\bf 47},  71 (1986).

\bibitem{sgmate2}
%%E. Vincent, J. Hammann, and M. Alba,
%%Solid State Comm. {\bf 58}, 57 (1986);
E. Vincent, and J. Hammann,
J. Phys. C{\bf 20}, 2659 (1987).

\bibitem{hukuprivate}
K. Hukushima and H. Kawamura, private communications.
\end  {thebibliography}
\end{document}